\documentclass[12pt]{article}
\usepackage{epsfig}
\usepackage{enumerate}
\usepackage{amsmath, amsfonts, amssymb}
\newcommand{\ct}[1]{#1}
\renewcommand{\ct}[1]{ \cite{#1}}

\newcommand{\bes}{\mathcal{B}}
\newcommand{\acid}{\text{aa}}

\newcommand{\mon}{\begin{displaymath}}
\newcommand{\moff}{\end{displaymath}}
\newcommand{\eon}{\begin{equation}}
\newcommand{\eoff}{\end{equation}}
\newcommand{\eq}[1]{Eq. \ref{#1}}

\newenvironment{changemargin}[2]{%
 \begin{list}{}{%
  \setlength{\topsep}{0pt}%
  \setlength{\leftmargin}{#1}%
  \setlength{\rightmargin}{#2}%
  \setlength{\listparindent}{\parindent}%
  \setlength{\itemindent}{\parindent}%
  \setlength{\parsep}{\parskip}%
 }%
\item[]}{\end{list}}

\long\def\symbolfootnote[#1]#2{\begingroup%
\def\thefootnote{\fnsymbol{footnote}}\footnote[#1]{#2}\endgroup}
\renewcommand{\baselinestretch}{1.0}

\title{Synonymous codon usage and\\
selection on proteins}
\date{October 13, 2004}

\author{Joshua B. Plotkin$^1$, Jonathan Dushoff$^2$, Michael M.
Desai$^3$,
Hunter B. Fraser$^4$}
\begin{document}

\maketitle

\begin{center} $^1$\textsc{Harvard Society of Fellows\\ 
and Bauer Center for Genomics Research \\
7 Divinity Avenue, Cambridge MA 02138, USA} \end{center} 

\begin{center} $^2$\textsc{Department of Ecology and Evolutionary Biology\\ 
Princeton University, Princeton, NJ 08540, USA} \end{center} 

\begin{center} $^3$\textsc{Department of Physics \\
and Department of
Molecular and Cellular Biology\\ 
Harvard University, Cambridge, MA 02138, USA} \end{center} 

\begin{center} $^4$\textsc{Department of Molecular and Cell Biology\\
University of California, Berkeley, CA, 94720, USA} \end{center}

\bigskip

\vspace*{.3in}

\pagebreak

\begin{changemargin}{.9cm}{.9cm}

\noindent Selection pressures on proteins are usually measured by
comparing homologous nucleotide sequences\ct{ZuckPaul65}.  Recently we
introduced a novel method, termed `volatility', to estimate selection
pressures on protein sequences from their synonymous codon
usage\ct{PlotDush03,PlotDush04}.  Here we provide a theoretical
foundation for this approach.  We derive the expected frequencies of
synonymous codons as a function of the strength of selection, the
mutation rate, and the effective population size.  We analyze the
conditions under which we can expect to draw inferences from biased
codon usage, and we estimate the time scales required to establish and
maintain such a signal.  Our results indicate that, over a broad range
of parameters, synonymous codon usage can reliably distinguish between
negative selection, positive selection, and neutrality.  While the power
of volatility to detect negative selection depends on the population
size, there is no such dependence for the detection of positive
selection.  Furthermore, we show that phenomena such as transient
hyper-mutators in microbes can improve the power of volatility to detect
negative selection, even when the typical observed neutral site
heterozygosity is low.

\end{changemargin}

\section{Introduction}

Nucleotide coding sequences of many organisms exhibit significant codon
bias -- that is, unequal usage of synonymous codons.  Codon bias has
been attributed both to neutral processes, such as asymmetric mutation
rates, as well as to selection acting on the synonymous codons
themselves. The most common selective explanation of codon bias posits
that synonymous codons differ in their fitness according to the relative
abundances of iso-accepting tRNAs; a codon corresponding to a more
abundant tRNA would be used preferentially so as to increase
translational efficiency\ct{Ikem81,DebrMarz94,SoreKurl89}.  To a large
extent, this hypothesis has successfully
explained interspecific variation in genome-wide codon usage for
organisms ranging from \textit{Escherichia coli} to \textit{Drosophila
melanogaster}\ct{Akas01}.

Recently, however, we have noted that codon bias in a protein sequence
can also result from selection at the amino acid level, even in the
absence of direct selection on synonymous codons
themselves\ct{PlotDush03,PlotDush04}. Codon bias arises from selection
at the amino acid level because of asymmetries in the structure of the
standard genetic code. Proteins that experience different selective
regimes should exhibit different synonymous codon usage.  Following from
this observation, we have introduced methods to screen a single genome
sequence for estimates of the selection pressures acting on its proteins
by comparing their synonymous codon usage\ct{PlotDush04}.

In this paper, we provide a theoretical discussion of codon usage biases
that result from selection at the amino acid level.  Our analysis helps
to provide a theoretical grounding for techniques of estimating
selection pressures on proteins using signals gathered from their
synonymous codon usage\ct{PlotDush03,PlotDush04}. Throughout most of
this paper, we will ignore any source of direct selection on synonymous
codons, and focus on the codon biases that result purely from selection
at the amino acid level.  To the extent that any other sources of codon
bias apply equally across the genome, we have devised a bootstrap method
to control for these external sources of codon bias when estimating
selection pressures on proteins\ct{PlotDush03,PlotDush04}.  In the
discussion, however, we describe a range of confounding factors that may
vary across the genome in some organisms and limit the applicability of
codon-based methods to detect selection.

\section{Codon volatility}

Codon usage biases can arise from the familiar process of
selection on proteins because synonymous codons may differ in their
\textit{volatility} -- defined, loosely, as the proportion of a codon's
point mutations that result in an amino acid
substitution\ct{PlotDush03}. Although there are several possible
definitions of volatility, which can all be informative, we have
recently used the following formal definition\ct{PlotDush04}.

We index the 61 sense codons in an arbitrary order $i=1\ldots61$.  We
use the notation $\acid(i)$ to denote the amino acid encoded by codon
$i$.  For each codon $i$, let $B(i)$ denote the set of sense codons
that differ from codon $i$ by a single point mutation. 
We define the volatility of codon $i$ by:
\begin{equation}
\nu(i) = \frac{1}{\#B(i)}\sum_{j \in B(i)}
D[\acid(i),\acid(j)]\\
\label{voldef}
\end{equation}
where $D$ denotes the Hamming metric, which is zero if two amino acids
are identical, and one otherwise.  The definition in Eq. ~\ref{voldef}
applies when all nucleotide mutations occur at the same rate.
When differential nucleotide mutation rates are known
(\textit{e.g.} a transition/transversion bias\ct{Wake96}), these rates can be
incorporated into the definition of codon volatility by appropriately
weighting the ancestor codons\ct{PlotDush04}.

Minor variants of Eq. \ref{voldef} yield related definitions of codon
volatility. For some applications, one may want to allow termination
codons in the definition of $B(i)$. It is also natural to consider
alternatives to the Hamming metric, $D$, that weight substitutions
between amino acids depending upon the differences in their
stereochemical properties\ct{MiyaMiya79,PlotDush03}. A variety of other
metrics\ct{TangWyck04,YampStol04} that reflect the effects of different
amino acid substitutions on protein structure may likewise be
incorporated into the definition of codon volatility.  In this paper,
however, we will focus on the most basic definition of codon volatility
(Eq.  \ref{voldef}, using the Hamming metric), because variant
definitions are based on the same underlying principle and produce
similar results in practice\ct{PlotDush03}.

Under the most basic definition of volatility, there are four amino
acids (Glycine, Leucine, Arginine, and Serine) whose codons differ in
their volatility.  As a result, when controlling for amino acid content,
we obtain a volatility signal from only those sites that contain one of
these four amino acids -- which amounts to about 30\% of the sites in a
typical gene. (If one uses stereochemical
metrics\ct{MiyaMiya79,PlotDush03} for $D$ in the definition of
volatility, then $\sim\!75$\% of the sites in a gene contain a
volatility signal).  Although 30\% may seem like a small proportion of
sites from which to obtain a signal of selective pressures, it is larger
than the proportion of sites often used to detect selection via sequence
comparison of recently diverged species\ct{FleiAlla02,ClarkGlan03}. (For
example, fewer than 4\% of neutral sites exhibit substitutions when
comparing human and chimpanzee sequences\ct{ClarkGlan03}.)

In the following sections we analyze the consequences of selection on
proteins for codon usage in general, as well as for the volatility
measure in particular.  We demonstrate that the expected codon usage at
a site, as well as its temporal dynamics, depend upon the strength of
positive or negative selection on the amino acid sequence.  In Sections
\ref{QSmodel} through \ref{PopSize} we examine negative selection in
infinite and finite populations. In Section \ref{PosSel} we discuss
positive selection.  Our analysis is initially confined to the patterns
of codon usage at a single site under selection at the amino acid level.
Proceeding from this analysis, we also discuss codon usage over many
sites within a gene or genome, and analyze how many sites are required
in principle to detect a reliable signal of selection by inspecting
synonymous codon usage. 

\section{Negative Selection and Codon Bias in an Infinite Population}

\label{QSmodel}

Most nonsynonymous mutations in a protein coding sequence presumably
reduce the fitness of an organism. For a large proportion of sites,
therefore, natural selection opposes any change in the amino acid.  We
refer to this type of selection as ``negative selection.''  

For the purposes of exploring the effect of negative selection on codon
usage, we assume that selection cannot discriminate between the
synonymous codons for the favored amino acid at a site.
However, mutations are more likely to be nonsynonymous, and hence
deleterious, if the codon at that site has high volatility. As we will
show, this fact results in an effective preference for the less
volatile codons, among those codons that code for the favored amino acid
at the site. We emphasize that this preference for a codon of low
volatility at a site under negative selection is \textit{not} caused by
a direct fitness difference between synonyms.  Rather, more volatile
codons will occur less frequently as a second-order consequence of
negative selection at the amino acid level, and the structure of the
genetic code.

Proteins with a larger number of sites under negative selection will
exhibit a statistical bias towards less volatile codons, after
controlling for their amino acid content.  
Here we calculate the expected magnitude of the codon bias as a function
of the mutation rate, the strength of negative selection, and, in
Section \ref{Wright}, the population size.  We also analyze the
conditions under which we can expect to detect and draw inferences from
this bias, and we estimate the time scales needed to establish and
maintain such a signal.

\subsection{A simplified genetic code}

In an infinite population, we can describe the dynamics of codon usage
at an individual site by using the standard multi-allele model first
introduced by Haldane\ct{Hald27} and used throughout the literature
(\textit{e.g.} ref.\ct{Nagy92} Eq. 2.25 or ref.\ct{Higg94}).  This model
describes a single site which can assume any of $K$ states. In order to
investigate codon usage, we consider $K = 64$ states, corresponding to
each of the $64$ possible codons.  In continuous time, the frequency
$x_i$ of individuals with codon $i$ evolves according to
\begin{equation} \frac{dx_i}{dt}=\sum_{j=1}^Kx_j(t) w_j M_{ij} - x_iW(t)
\label{QS} \end{equation} where $w_j$ is the Malthusian fitness of codon
$j$, $W(t) \equiv \sum_j w_j x_j(t)$ is the mean fitness of the
population, and $M_{ij}$ is the instantaneous rate of mutation from
codon $j$ to codon $i$, with $\sum_j M_{ij}=0$.  Although Eq. \ref{QS} is non-linear, the
equilibrium frequencies of the ``alleles" $i=1,2,\ldots K$ are given by
the leading eigenvector of the matrix $w_j M_{ij}$\ct{ThomMcBr74}.
These frequencies determine the expected equilibrium codon usage at a
site. For the purposes of this paper, alternative formulations of the
$K$-allele model that treat the processes of selection and mutation
separately (\textit{e.g.} ref\ct{CrowKimu70} Eq. 6.4.1) yield the exact same
results.

The equilibrium solution to Eq. \ref{QS} for the full genetic code does
not lend itself to intuitive understanding. Transient dynamics are also
difficult to calculate in this high-dimensional system.  Therefore, in
order to highlight the essential points of our analysis, we first
consider a ``toy'' genetic code that retains those features of the true
genetic code relevant to the study of synonymous codon usage under
negative selection.  As we will demonstrate, the solution for the
simplified genetic code yields a complete understanding for the full
genetic code as well.

We imagine a simplified genetic system with only three possible codons,
$a_1$, $a_2$, and $b$.  Codons $a_1$ and $a_2$ code for amino acid $A$,
which is favored, and codon $b$ encodes amino acid $B$, which has
selective disadvantage $\sigma$.  We assume that mutations occur at rate
$u$ between these codons according to the structure \[ a_1
\rightleftarrows a_2 \rightleftarrows b, \] so that of the two
synonymous codons, $a_2$ is more volatile.

According to the standard multi-allele model (\eq{QS}), the relative
frequencies of codons $a_1$, $a_2$, and $b$ are described by the
equation
\begin{equation} \label{trit}
\frac{d}{dt}\left (\begin{array}{c} a_1(t) \\ a_2(t) \\ b(t)
\end{array}\right ) = \left ( \begin{array}{ccc}
1-u & u & 0 \\ u & 1-2u & u(1-\sigma) \\ 0 & u & (1-u)(1-\sigma)
\end{array} \right ) \left (\begin{array}{c} a_1(t) \\ a_2(t) \\
b(t) \end{array}\right ) - W(t) \left (\begin{array}{c} a_1(t) \\
a_2(t) \\ b(t) \end{array}\right ),
\end{equation} 
where $W(t)=a_1(t)+a_2(t)+(1-\sigma)b(t)$. 

The equilibrium frequencies
of codons are given by the leading eigenvector of the matrix in Eq.
\ref{trit}. A simple perturbation analysis of this eigenvector shows that the
equilibrium frequency of $a_1$ depends monotonically
on $\sigma$, and it exhibits a sharp transition between two regimes: the
weak selection regime $\sigma \ll u$ and the strong selection regime
$\sigma \gg u$. In the weak selection regime, the equilibrium relative
frequencies of synonyms are given by the expansion
\eon \frac{\hat{a_1}}{\hat{a_1}+\hat{a_2}}=\frac{1}{2}+\frac{1}{12}
\frac{\sigma}{u} +
O\left(\frac{\sigma^2}{u^2}\right). \eoff
And in the strong selection regime, the equilibrium relative
frequencies are given by
\eon \frac{\hat{a_1}}{\hat{a_1}+\hat{a_2}}=\frac{\sqrt 5 -1}{2}-\frac{
(5-2\sqrt5)(1-\sigma)}{5} \frac{u}{\sigma} +
O\left(\frac{u^2}{\sigma^2}\right). \eoff

In the absence of selection $(\sigma=0)$ all three codons occur with
equal frequency, as we would expect.  In particular, the relative
frequency of the two synonymous codons $a_1$ and $a_2$ equals
$\frac{1}{2}$, regardless of the mutation rate.  For weak selection
($\sigma \ll u$), this result is still approximately true, according to
the perturbation expansion above.  In the case of strong negative
selection ($\sigma \gg u$), the relative frequency of the two synonymous
codons is given approximately by the inverse of the golden mean,
$\frac{\sqrt 5 -1}{2} \approx 0.62$. 

\begin{figure}[ht] \begin{center}
\epsfig{file=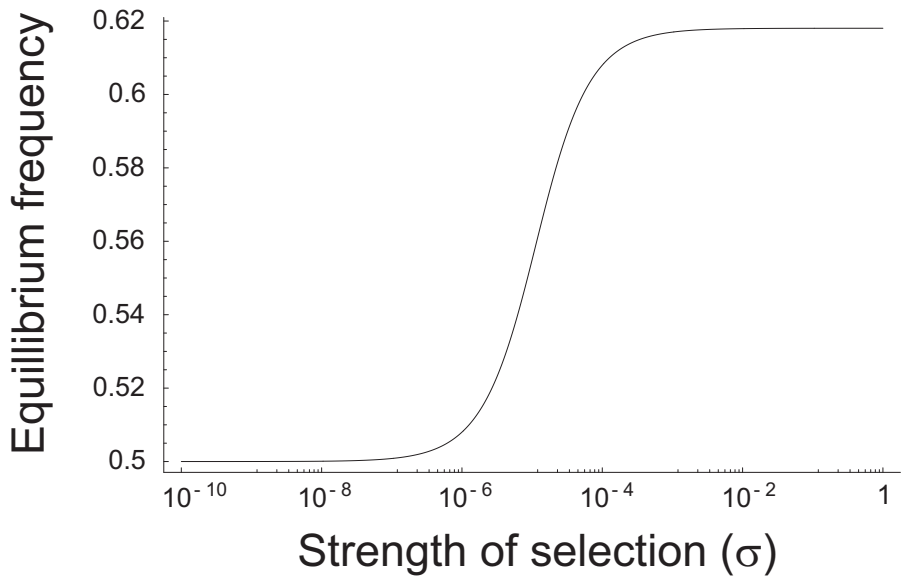,angle=0,width=12cm} \caption{The relationship
between selection at the amino acid level and resulting synonymous codon
usage. The graph shows relative equilibrium frequency of synonymous
codons,  $\hat{a_1}/(\hat{a_1}+\hat{a_2})$, as a function of the strength
of negative selection, $\sigma$. The relative frequency of codon $a_1$
is approximately $\frac{1}{2}$ in the weak selection regime ($\sigma \ll
u$), and approximately $\frac{\sqrt{5}-1}{2}$ in the strong selection
regime ($\sigma \gg u$). In this figure $u=10^{-5}$.} \label{TritEquil}
\end{center} \end{figure}

The sharp transition between the weak and strong selection regimes
defines $\sigma = u$ as a critical value for negative selection.  For
$\sigma \ll u$ negative selection is ineffective at favoring the less
volatile codon, and the site is effectively neutral.  But when $\sigma
\gg u$, negative selection favors the less volatile codon, and the
magnitude of this effect depends only weakly on the value of $\sigma$.
This is an essential point.  In the strong selection regime, the
magnitude of negative selection is relatively unimportant; volatile
codons are disfavored at all sites where $\sigma \gg u$. The transition
between the weak and strong selection regimes is shown in Fig.
\ref{TritEquil}.

\subsection{The effective disadvantage of a volatile codon}

The critical value of $\sigma$ discussed above can be understood
intuitively by considering the ``effective selective disadvantage" of
the more volatile codon $a_2$ that results indirectly from its
volatility.  We will use the notion of an ``effective selective
disadvantage" to aid in our analysis of codon usage at a
site under negative selection. But we emphasize that our model (Eq.
\ref{QS}) does not assume any direct fitness difference
between synonymous codons.

When the disfavored amino acid $B$ is lethal to the organism, then the
effective selective disadvantage of codon $a_2$ is particularly simple
to understand.  In this case, individuals with codon $a_2$ are removed
from the population at rate $u$ because they mutate to the lethal codon
$b$, but receive no back-mutations.  Hence the effective selective
disadvantage, denoted $s$, of codon $a_2$ versus codon $a_1$ is given by
$s = u$. The effective selective disadvantage of $a_2$ does not arise
from a fitness difference between synonyms, but rather from selection at
the level of amino acids and the structure of the genetic code.

When amino acid $B$ is not lethal the situation is slightly more
complicated.  Nevertheless, for $\sigma \gg u$, mutations from $a_2$ to
$b$ typically die due to negative selection before they mutate back from
$b$ to $a_2$. As a result, the effective selective disadvantage will
still be $s=u$ in the regime of strong selection. We can make this
argument concrete by considering the mutation-selection balance between
codon $b$ and codon $a_2$. According to the standard mutation-selection
balance, the equilibrium frequency of codon $b$ relative to codon $a_2$
equals $\frac{u}{\sigma}$ in the regime $\sigma \gg u$.  Thus for each
mutant from $a_2$ to $b$, there are at most of order $\frac{u}{\sigma}$
mutations from $b$ to $a_2$. The net mutation rate from $a_2$ to $b$ is
therefore $u \left(1-\frac{u}{\sigma} \right)$.  This is the rate at
which individuals of type $a_2$ are lost from the population due to the
fact that $a_2$ is more volatile than $a_1$.  Thus the effective
selective disadvantage of $a_2$ relative to $a_1$ is given by $s = u
\left( 1 - \frac{u}{\sigma} \right)$.  By definition, in the strong
selection regime we neglect $\frac{u}{\sigma}$ compared to 1, and the
effective selective disadvantage of codon $a_2$ is simply $s = u$.

A similar argument holds for the real genetic code.  In this case, the
favored amino acid may correspond to several synonymous codons, each
with a potentially different volatility. However, the effective
selective disadvantage, $s$, of a more volatile codon relative to a less
volatile synonym is simply the difference in the number of mutations
leading to a disfavored codon ($\sigma \gg u$) times $\frac{u}{3}$,
where $u$ is the nucleotide mutation rate. (Note that $\frac{u}{3}$ is
the rate of mutation between any two particular nucleotides.)   For
example, when considering the relative frequencies of codons AGA and CGG
at a site under negative selection for Arginine, AGA has selective
disadvantage $s=\frac{2}{3}u$ compared to CGG, since AGA has two more
disfavored neighbors than CGG. By using the value of the effective
selective disadvantage, $s$, we can calculate the equilibrium relative
frequency of any pair of synonymous codons in mutation-selection
balance, and thereby deduce the relative frequencies of all synonyms.
Therefore, we can predict synonymous codon usage in the genetic code
without resorting to the full solution of \eq{QS}.

An analogous argument can be used to calculate the effective selective
disadvantage of codon $a_2$ in the regime of weak selection ($\sigma \ll
u$). In this regime, the relative equilibrium frequency of codon $b$
versus codon $a_2$ equals $1-\frac{\sigma}{2u}$. Thus, the effective
selective disadvantage of $a_2$ versus $a_1$ is approximately $s = 0$,
plus a small correction of order $\sigma$.  In other words, when $\sigma
\ll u$ selection between $a_1$ and $a_2$ is effectively neutral; it
cannot generate codon bias.  We therefore refer to the regime
$\sigma \ll u$ as the ``almost neutral regime." This result holds both
for the simplified three-codon model and for the real genetic code.  

It is also important to calculate the amount of time required to reach
equilibrium codon usage in the presence of strong negative selection.
Explicit solution of Eq. \ref{trit}, assuming $\sigma \gg u$, indicates
that the $e$-fold relaxation time is of order $\frac{1}{u}$ (the
selection coefficient is $s \sim u$, and so the time scale for population
sizes to change under selection is of order $\frac{1}{s} \sim
\frac{1}{u}$).  In other words, starting from any initial frequencies
$a_1(0)$ and $a_2(0)$, these frequencies will become $e$-fold closer to
their equilibrium values after a duration of order $\frac{1}{u}$
generations.  The same time scale holds for almost neutral sites
($\sigma \ll u$) and for the real genetic code\symbolfootnote[2]{For
$\sigma \ll u$, the process is almost neutral and the time scale
calculation of Section \ref{relax} applies.  The real genetic code has
the same dynamics because we still have $s \sim u$ for $\sigma \gg u$
and neutral behavior for $\sigma \ll u$.}.  In practice, $u$ will
be quite small, and equilibrium volatility is approached very slowly.
We will revisit this point when we discuss finite populations, and again
when we discuss positive selection.

\subsection{A specific example: selection for Arginine}

In this section we consider a simple example that demonstrates how
our analysis applies to the real genetic code. We use Eq. \ref{QS} to
model the dynamics of $K=64$ alleles corresponding to the 64 codons,
indexed in an arbitrary order. For our example, we consider a single
site under negative selection for an Arginine codon. In this case we
define
\begin{equation}
M_{ij}=\begin{cases}
1-3u, \text{\ \ \ if $i$=$j$}\\
u/3, \text{\ \ \ if $i$ and $j$ differ by a point mutation} \\
0, \text{\ \ \ otherwise}
\end{cases}
\end{equation}
where $u$ is the nucleotide mutation rate. We define
\begin{equation}
w_i=\begin{cases}
1, \text{\ \ \ if $i$ encodes Arginine} \\
1-\sigma, \text{\ \ \ if $i$ encodes a non-Arginine amino acid}\\
1-\gamma, \text{\ \ \ if $i$ encodes stop}
\end{cases}
\end{equation}
so that a codon encoding an amino acid other than Arginine has
fitness $1-\sigma$, and a termination codon has fitness
$1-\gamma$.  We analyze this model numerically by calculating the
leading eigenvector of the matrix $w_j M_{ij}$, which yields the
equilibrium frequencies of all 64 codons.

In the case of no selection ($\sigma = \gamma = 0$), we find that all
codons occur with the same equilibrium frequency, independent of
mutation rate, as we would expect.  For almost neutral selection
($\sigma \sim \gamma \ll u$), codon usage is still approximately
uniform. In the opposite case when Arginine is favored and all other
amino acids (or termination codons) are strongly disfavored
(\textit{i.e.} $\sigma \sim \gamma \gg u$), the Arginine codons CGA,
CGG, CGC, CGT, AGA, and AGG occur with equilibrium relative frequencies
$\approx$ 0.214 : 0.214 : 0.191 : 0.191 : 0.095 : 0.095.  As expected,
under negative selection the more volatile Arginine codons occur with
lower relative frequency in equilibrium.

\begin{figure}[ht] \begin{center}
\epsfig{file=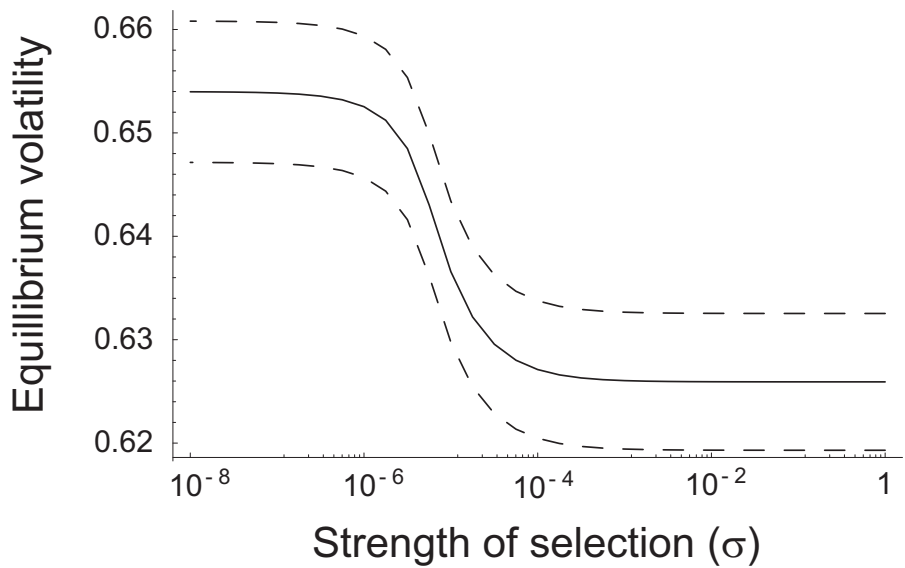,angle=0,width=12cm} \caption{The relationship
between selection and volatility for a gene comprised of $L=200$ freely
recombining sites under selection for Arginine. The graph shows expected
volatility per site in the gene ($\pm 1$ standard deviation, dashed) as
a function of the strength of negative selection, $\sigma$. The
nucleotide mutation rate is $u=10^{-5}$.  The expected volatility is
significantly depressed in the regime of strong negative selection,
$\sigma \gg u$.  (For this figure we assume $\gamma = 1$; virtually
identical results hold for $\gamma = \sigma$.) } \label{argex}
\end{center} \end{figure}

The equilibrium frequencies of Arginine codons determine the expected
volatility at a single Arginine site under negative selection.
Assuming free recombination\ct{SawyHart92}, an individual gene consists
of many such sites randomly assembled; the mean and standard deviation
in the volatility (per site) of a randomly sampled gene are shown in
Fig.  \ref{argex}, as a function of the strength of negative selection
$\sigma$.  Note that the stronger the negative selection, the lower the
expected equilibrium volatility. The expected volatility exhibits a
sharp transition from high to low values when the strength of negative
selection $\sigma$ reaches the mutation rate $u$, as discussed above.
On either side of this transition, the volatility is insensitive to
$\sigma$. The standard deviations plotted in Fig. \ref{argex} correspond
to a gene comprised of $L=200$ such sites, each modeled independently by
the multi-allele equation.

According to Fig.  \ref{argex}, $L=200$ independent sites that each
experience neutrality ($\sigma \ll u$) can be distinguished on the basis
of their volatility from $L=200$ sites that experience negative
selection ($\sigma \gg u$).  The difference in the expected volatility
between these two regimes is greater than four standard deviations of
the volatility within either regime.

In reality, the selective constraint $\sigma$ will vary greatly across
the sites of a given protein.  In this case, disregarding the
possibility of positive selection, the volatility of a gene (after
controlling for its amino acid sequence) essentially reflects the
relative number of informative sites that experience negative selection
versus neutrality.  For example, the volatility of gene $X$ that
contains $L=200$ informative sites under negative selection and an equal
number of neutral sites will be significantly greater (with a
$Z$-score of about three) than the volatility of gene $Y$ that
consists of $2L$ informative sites all under negative selection.
A more thorough discussion of variable selection pressures across genes
is described in Section \ref{Infer}, below.

Table \ref{GLRS} shows the equilibrium relative frequencies of
synonymous codons for each of the informative amino acids (G, L, R, and
S) under neutrality versus various selective regimes.  In Table
\ref{GLRS} we assume, as we do throughout this manuscript, that
volatility is measured using the Hamming metric and that there is no
transition/transversion bias.  Corresponding values for different
metrics or including a mutational bias may be calculated using the same
approach. As seen in Table \ref{GLRS}, the difference in the expected
volatility between selective regimes is least extreme (indeed, barely
informative) for Glycine sites.  The volatility difference is most
extreme for serine sites: the highly volatile codons AGT and AGC are not
expected to occur at a site under negative selection, but they
preferentially occur at a site under positive selection.  This extreme
case results from the fact that codons AGT and AGC are not connected by
synonymous point mutations to the other serine codons. 
This situation does not imply that codons AGT and AGC should be treated
separately from the other serine codons. In fact, when treated as an
entire group, the serine codons are particularly informative for
positive selection (Table \ref{GLRS}).

\renewcommand{\baselinestretch}{.9}
\begin{table}
{\small
\begin{center}
\begin{tabular}{lllllc}
& Neutral & Neutral* & Negative & Positive & $\nu$\\
\textbf{Leucine} & & & & &\\
cta & 0.16667 & 0.17300 & 0.21353 & 0.14213 & 5/9\\
ctc & 0.16667 & 0.18580 & 0.19098 & 0.17056 & 6/9\\
ctg & 0.16667 & 0.17890 & 0.21353 & 0.14213 & 5/9\\
ctt & 0.16667 & 0.18580 & 0.19098 & 0.17056 & 6/9 \\
tta & 0.16667 & 0.12990 & 0.09549 & 0.18274 & 5/7\\
ttg & 0.16667 & 0.14650 & 0.09549 & 0.19188 & 6/8 \\
\hline $\mathbb{E}[\nu]$ & 0.65146 & 0.64590 & 0.63172 & 0.65978 &\\
$\sigma[\nu]$ & 0.07362 & 0.07259 & 0.07022 & 0.07217 &\\
\\
\textbf{Arginine} & & & & &\\
aga & 0.16667 & 0.15210 & 0.09549 & 0.19149 & 6/8\\
agg & 0.16667 & 0.17050 & 0.09549 & 0.19859 & 7/9\\
cga & 0.16667 & 0.15210 & 0.21353 & 0.12766 & 4/8\\
cgc & 0.16667 & 0.17740 & 0.19098 & 0.17021 & 6/9\\
cgg & 0.16667 & 0.17050 & 0.21353 & 0.14184 & 5/9\\
cgt & 0.16667 & 0.17740 & 0.19098 & 0.17021 & 6/9\\
\hline $\mathbb{E}[\nu]$ & 0.65278 & 0.65400 & 0.62592 & 0.66766 & \\
$\sigma[\nu]$ & 0.09854 & 0.09660 & 0.09354 & 0.09528 & \\
\\
\textbf{Serine} & & & & & \\
agc & 0.16667 & 0.18510 & 0.00000 & 0.20636 & 8/9\\
agt & 0.16667 & 0.18510 & 0.00000 & 0.20636 & 8/9\\
tca & 0.16667 & 0.13440 & 0.25000 & 0.13265 & 4/7\\
tcc & 0.16667 & 0.17190 & 0.25000 & 0.15477 & 6/9\\
tcg & 0.16667 & 0.15162 & 0.25000 & 0.14510 & 5/8\\
tct & 0.16667 & 0.17190 & 0.25000 & 0.15477 & 6/9\\
\hline $\mathbb{E}[\nu]$ & 0.71792 & 0.72981 & 0.63243 & 0.73970 & \\
$\sigma[\nu]$ & 0.12504 & 0.12561 & 0.03913 & 0.12847 & \\
\\
\textbf{Glycine} & & & & & \\
gga & 0.25000 & 0.22460 & 0.25000 & 0.23810 & 5/8\\
ggc & 0.25000 & 0.26180 & 0.25000 & 0.25397 & 6/9\\
ggg & 0.25000 & 0.25170 & 0.25000 & 0.25397 & 6/9\\
ggt & 0.25000 & 0.26180 & 0.25000 & 0.25397 & 6/9\\
\hline $\mathbb{E}[\nu]$ & 0.65625 & 0.65724 & 0.65625 & 0.65675 & \\
$\sigma[\nu]$ & 0.01804 & 0.01859 & 0.01804 & 0.01775 & \\
\end{tabular}
\end{center}
\caption{Equilibrium codon usage under neutrality versus selective
regimes.  In each selective regime, we report the equilibrium relative
abundance of codons, and the resulting mean and standard deviation in
volatility per site. The first column corresponds to neutrality
($\sigma=\gamma \ll u$); the second column corresponds to neutrality but
with disfavored termination codons ($\sigma \ll u$, $\gamma=1$); the third
column corresponds to strong negative selection in an infinite
population ($\sigma \gg u$, $\gamma \gg u$); the fourth column
corresponds to the expected frequencies after a positively selected
sweep (see Section \ref{PosSel}). The final column gives the volatility
of each codon, assuming no transition/transversion bias\ct{PlotDush04}.}
\label{GLRS} } \end{table}
\renewcommand{\baselinestretch}{1.0}

\section{Negative Selection in a Finite Population} \label{Wright}

The models presented in Section \ref{QSmodel} describe the processes of
mutation and negative selection in an infinite population. In finite
populations, however, genetic drift also affects allelic frequencies.
In this section, we study the combined effects of mutation, negative
selection, and drift, which we analyze using diffusion equations.  These
equations can be very complex.  A full treatment of even the simplified
three-codon genetic code requires a two-dimensional diffusion process,
and the real genetic code involves a $63$-dimensional process.  To make
this problem tractable, we use the notion of the ``effective selective
disadvantage" of more volatile codons, as discussed above.  This allows
us to consider the dynamics only at the favored codons, thereby reducing
the dimensionality of the diffusion process.

The neutral ($\sigma = 0$) or almost neutral ($\sigma \ll u$) regimes
are straightforward: here all synonymous codons for the favored amino
acid have the same effective fitness.  In this regime, each synonymous
codon occurs with the same probability in steady state, independent of
population size.  

For the remainder of this section, we analyze the case of strong
negative selection ($\sigma \gg u$) at a single site.   We consider a
diffusion approximation to the process of mutation, selection, and drift
operating only on the synonymous codons, to each of which we assign an
effective selective coefficient. For the simplified three-codon
genetic system, the more volatile codon $a_2$ has an effective selective
disadvantage of $s = u$ compared to codon $a_1$.  For the real genetic
code, more volatile codons will have a selective disadvantage of this
order, but the precise value of $s$ will depend on the specific amino
acid in question.  In the following analysis, we consider the case of
the simplified three-codon system.  However, we do not explicitly make
the substitution $s = u$, so that our results can also be applied (with
a slightly different value of $s$) to the real genetic code.

The time-dependent frequency $f(x,t)$ of allele $a_1$ relative to
allele $a_2$ can be described by the Komolgorov forward
equation\ct{KimuCrow64}
\begin{equation} \frac{\partial f(x,t)}{\partial t} =
-\frac{\partial}{\partial x} \{a(x) f(x,t)\} +
\frac{1}{2}\frac{\partial^2}{\partial x^2} \{b(x)f(x,t)\} \end{equation}
where the instantaneous mean and variance in the change of allelic
frequency are given by
\begin{eqnarray*}
a(x)&=&sx(1-x)-ux+u(1-x)\\
b(x)&=&x(1-x)/N.
\end{eqnarray*}
The stationary distribution of allele frequencies $\hat f(x)$ satisfies
the equation
\begin{equation}
\frac{d}{dx}\{b(x)\hat f(x)\}=2 a(x)\hat f(x)
\end{equation}
which has the solution\ct{Wrig31} \eon \hat
f(x)=Cx^{\theta-1}(1-x)^{\theta -1} \ e^{S x} \eoff where $\theta=2Nu$,
$S=2Ns$, and $C$ is chosen so that $\int_0^1\hat f(x)  dx=1.$  Since $s
\sim u$ (and thus $S \sim \theta$), the shape of the stationary the
distribution $\hat f(x)$ falls into two categories: a bell-shaped
distribution in the regime $\theta>1$, and a U-shaped distribution in
the regime $\theta<1$. In other words, for $\theta>1$ the steady-state
population is typically polymorphic at the locus, much like the infinite
population mutation-selection balance.  Whereas for $\theta<1$ the
steady-state population is usually near-monomorphic at the locus,
occasionally switching between alleles $a_1$ and $a_2$, with a bias
(whose strength is determined by $S$) towards allele $a_1$ .

In stationary state, 
the expected frequency of allele $a_1$ is given by
\begin{equation}
M(\theta,S)=\int_0^1x \hat f(x) dx =\frac{1}{2}+\frac{\bes(\theta+1/2,
S/2)}{2\bes(\theta-1/2,S/2)}
\label{mean}
\end{equation}
where $\bes(x,y)$ is the modified Bessel function of the first kind.
Similarly, the variance in the frequency of allele $a_1$ is given by
\begin{eqnarray}
V(\theta,S)&=&\int_0^1 x^2 \hat f(x) dx - M(\theta)^2\\
&=&\frac{1}{4+8\theta}+\frac{2\theta \bes(\theta-1/2,S/2)
\bes(\theta+3/2,S/2)-(1+2\theta)
\bes(\theta+1/2,S/2)^2}{(4+8\theta)\bes(\theta-1/2,S/2)^2}
\label{var}
\end{eqnarray}
We use the standard Taylor series expansion of
$\bes(x,y)$, 
\begin{equation}
\bes(x,y)=\sum_{m=0}^{\infty}\frac{(y/2)^{x+2m}}{m!\Gamma(x+m+1)}, 
\label{Bexpand}
\end{equation} 
to obtain a simple approximation for the mean
stationary frequency of allele $a_1$:
\begin{equation}
M(\theta,S) = \frac{1}{2}+\frac{S}{4} + O(\theta^2),
\label{nearmean}
\end{equation}
valid for $\theta \sim S \ll 1$. This approximation indicates that the
difference in expected volatility at a site under neutral versus
negative selection is of order $S$, when $\theta \ll 1$. 

When $\theta=S=1$, the mean stationary frequency of allele $a_1$ assumes
the value $\frac{1}{e-1}\approx 0.58$. For $\theta \sim S \gg 1$, the
mean frequency quickly approaches the asymptotic value
$\lim_{\theta\rightarrow \infty} M(\theta,\theta)=\frac{\sqrt{5}-1}{2}$,
in agreement with our earlier result for an infinite population. 

The results in this section generalize our analysis of an infinite
population.  For an infinite population, we found that the expected
relative frequency of codon $a_1$ equals $\frac{1}{2}$ in the almost
neutral regime, and it equals $\frac{\sqrt 5 -1}{2}$ in the strong
selection regime. In a finite population with $\theta \gg 1$, the same
results hold. In a finite population with $\theta \ll 1$, the
expected relative frequency of the more volatile codon equals
$\frac{1}{2}$ in the neutral regime, and it equals
$\frac{1}{2}+\frac{Ns}{2}$ in the strong selection regime.  For any
population size, the relative frequency of codon $a_1$ depends
monotonically on the strength of selection at the amino acid level,
$\sigma$, and it exhibits a sharp transition at the critical value
$\sigma=u$.

It is worth noting that our exact expression (Eq. \ref{mean}) for the
mean stationary frequency of allele $a_1$ generalizes earlier work by
Bulmer\ct{Bulm91} on the relative frequency of two synonymous codons
that experience a direct fitness difference. In the limit of small
$\theta$, we find that
\begin{equation} \lim_{\theta\rightarrow0} M(\theta,S) = \frac{1}{2} +
\frac{\bes(1/2, S/2)}{2\bes(-1/2,S/2)} = \frac{1}{1+e^{-S}},
\end{equation}
which agrees with Bulmer's result (his Equation 6). In other words, Bulmer's
approximation applies only for vanishing small mutation rates (or
population sizes).

We can again use the standard Taylor expansion of the Bessel function to
obtain a simple expression for the variance in the stationary
frequency of allele $a_1$,
\begin{equation}
V(\theta,S) \approx
\frac{(3+2\theta)(4+8\theta)-3S^2}{16(3+2\theta)(1+2\theta)^2},
\label{nearvar}
\end{equation}
which is a highly accurate approximation for all $\theta$,
provided as usual that $S$ is of order $\theta$ or smaller. Note that
when $\theta \ll 1$ the variance is approximated by
$\frac{1}{4}-\frac{\theta}{2}$, and when $\theta \gg 1$ the variance is
of order $\frac{1}{\theta}$.

\subsection{Inferring Negative Selection in a Finite Population}
\label{Infer}
\begin{figure}[ht] \begin{center}
\epsfig{file=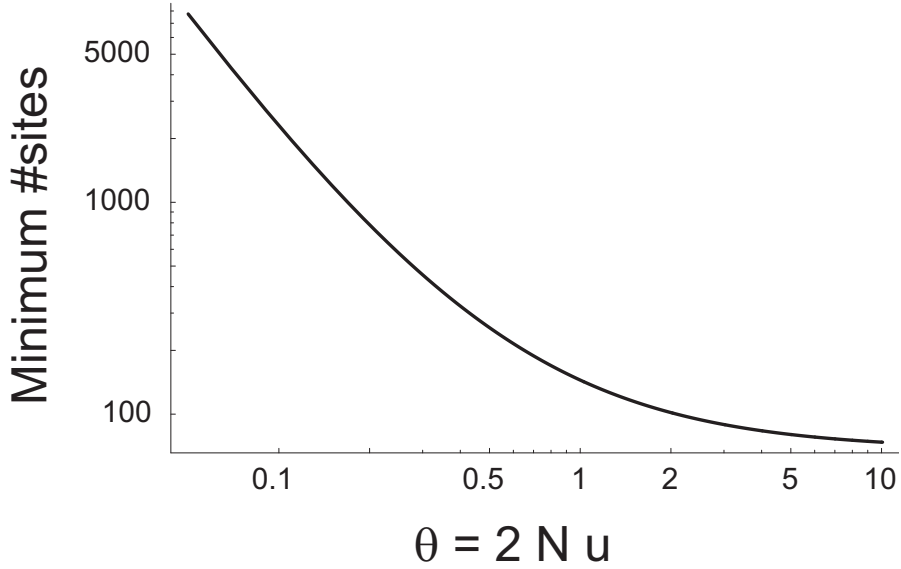,angle=0,width=12cm} \caption{The relationship
between the scaled population size, $\theta=2Nu$, and the minimum number
of sites required to distinguish negative selection from neutrality, at
the 95\% confidence level. Sites are assumed to be unlinked.  It is
important to note that the appropriate effective population size that
determines the value of $\theta$ in practice does not necessarily equal
the average neutral site heterozygosity (see Section \ref{PopSize}).}
\label{minL} \end{center} \end{figure}

Our exact (Eq. \ref{mean}) or approximate (Eq.  \ref{nearmean})
expressions for the stationary mean frequency of codon $a_1$ allow us to
determine the minimum number of sites required for codon volatility to
distinguish reliably between neutral versus negative selection.  When
sites are modeled independently (equivalent to the assumption of
linkage equilibrium\ct{SawyHart92}), under neutrality ($\sigma \ll u$;
$s=0$) the relative frequency of codon $a_1$ versus codon $a_2$ across a
gene of length $L$ is binomially distributed with mean $\frac{1}{2}$ and
variance $\frac{1}{4L}$.  If, on the other hand, the gene experiences
negative selection ($\sigma \gg u$; $s=u$), then the relative frequency
of codon $a_1$ is binomially distributed with mean $M(\theta,S)$ and
variance $M(\theta,S)[1-M(\theta,S)]/L$.  Therefore, in order to
reliability reject neutrality at about the 95\% confidence level, we
require \begin{equation} \label{minLeq} M(\theta,S)-\frac{1}{2} \ > \  2
\sqrt{\frac{1}{4L}} \end{equation} Using this equation, Fig. \ref{minL}
shows the minimum number of sites required to reliably distinguish
negative selection from neutrality on the basis of codon volatility,
under our simplified 'genetic code' consisting of three codons. 

Eq. \ref{minLeq} applies when comparing a collection of neutral
sites against a collection of sites under negative selection.  In most
situations, however, the selective constraint $\sigma$ will vary across
the sites of a protein. For example, consider gene $X$ with $L+J$ sites
under negative selection, compared to gene $Y$ with $L$ neutral sites
and $J$ sites under negative selection. In this case, the expected
frequency of codon $a_1$ in gene $Y$ is $(L/2 + J M(\theta,S))/(L+J)$.
Therefore, in order to reliably infer that gene $X$ experiences more
negative selection than gene $Y$, at the 95\% confidence level we
require \begin{equation} \label {minLeq2} M(\theta,S) - \frac{L/2 +
JM(\theta,S)}{(L+J)} \  > \ 2 \sqrt{\frac{L/4 + J M(\theta,S)
[1-M(\theta,S)]}{(L+J)^2}} \end{equation} As Eq. \ref{minLeq2}
indicates, the power to discriminate between two genes is decreased when
both genes contain many sites, $J$, under negative selection and only a
few sites, $L$, under different selective regimes. Nevertheless,
provided $J \sim L$, the power to discriminate between genes $X$ and $Y$
is decreased by $\sim$20\% at most (compared to $J=0$), and so the
minimum number of sites required to detect negative selection (Fig.
\ref{minL}) remains mostly unchanged.

Although the results in this section were derived for a simplified
genetic code, the scaling behavior of these solutions holds for the full
genetic code as well -- \textit{i.e.} when comparing neutrality to
negative selection, for $\theta \ll 1$ the expected difference in
volatility per site will be of order $\theta$; and for $\theta \gg 1$
the expected difference in volatility can be calculated from the
infinite population model (Eq. \ref{QS} and Table \ref{GLRS}). 

\subsection{Relaxation towards steady state} \label{relax} Although Eq.
\ref{mean} predicts the steady-state relative frequencies of codons
$a_1$ and $a_2$ in the selected regime ($\sigma \gg u$), we have not yet
discussed how long it takes, on average, to reach this steady
state. In the case of a very large population, $\theta \gg 1$, we know
from the infinite population model (Section \ref{QS}) that the $e$-fold
relaxation time to equilibrium is of order $\frac{1}{u}$ generations. In
this section, we demonstrate that the same result applies to the
time scale of relaxation towards steady state in the regime $\theta \ll
1$.

As usual, we consider a single site under negative selection. In the
regime $\theta \ll 1$, we have seen that the steady-state population
will spend most of the time in a nearly monomorphic state, with a
preference (of order $\theta$) for the less volatile codon, $a_1$.
Therefore, in order to calculate the time scale of relaxation towards
steady state, we may simply calculate the amount of time required such
that, starting with a population fixed for allele $a_2$, the probability
of the population remaining fixed for allele $a_2$ has been reduced
$e$-fold.

Given a population initially fixed for codon $a_2$, there are $Nu$
mutations to codon $a_1$ generated per generation. Each of these
mutations has an effective selective advantage $s=u$ over allele $a_2$,
and will therefore fix with probability
$2s/(1-e^{-2Ns})$\ct{CrowKimu70}. Hence the rate of production of a
mutation that will eventually fix is given by \begin{equation} P_{fix} =
\frac{2Nus}{1-e^{-2Ns}} \approx u, \label{Pfix} \end{equation} assuming
$\theta \ll 1$.  According to this calculation, the mean time until
fixation of codon $a_1$ is of order $\frac{1}{u}$ generations, which
gives the time scale of relaxation to the steady-state codon usage in a
finite population under negative selection.

\section{About Population Sizes} \label{PopSize} As discussed above, the
strength of the signal of negative selection depends upon the
parameter $\theta = 2Nu$. What is the appropriate value of
$\theta$ in practice?

Unfortunately, this question is far easier asked than answered.
Population geneticists have long struggled to reconcile estimates of
$\theta$ deduced from polymorphism data with direct measurements of $N$
and $u$ across broad taxonomic ranges.  The effective population sizes
of micro-organisms in particular are topics of active debate.  Estimates
of $\theta$ are usually obtained by comparing SNP data at neutral (or
presumably neutral) sites against the expected site diversity
or the expected number of segregating sites under a neutral
model\ct{Ewen04}. In a recent survey\ct{LyncCone03} authors have
reported an average value of $\theta \approx 0.15$ among the prokaryotes
studied. But estimates of $\theta$ for a microbial species can
vary by four orders of magnitude, and they depend strongly on
assumptions about population structure\ct{Berg96}.  To complicate
matters further, heterogeneity in mutation rates leads to substantial
underestimates of $\theta$\ct{Taji96}.  

Aside from uncertainty in its estimation, the value of $\theta$ deduced
from neutral SNP data\ct{LyncCone03} may not be relevant to questions of
selection and volatility.  Monomorphism observed at neutral sites may
result from non-neutral processes, such as background
selection\ct{CharMorg93} or hitchhiking on periodically sweeping
sites\ct{MaynHaig74}.  As a result, the variance effective population
size estimated from SNP data may not be relevant to other aspects of
evolution, such as substitutions at linked weakly selected
sites\ct{Gill01}.  

One particularly striking example of a discrepancy in the appropriate
effective population sizes arises from the consideration of mutator
phenotypes. Populations of microbial species periodically experience a
transient increase in the mutation rate, often $10^2-10^3$ times greater
than that of a non-mutator strain\ct{GiraRadm01}.  Between 2-20\% of
bacterial populations isolated in the wild at any given time exhibit a
mutator phenotype\ct{GiraRadm01,OlivCant00,LeclLi96}. The mutator phase
can be induced in several ways. A defective DNA repair gene may arise
and sweep to fixation by hitchhiking on a positively selected
mutation\ct{NotlSeet02}. The entire population then experiences an
elevated mutation rate until a non-mutator allele sweeps and replaces
the mutator\ct{NotlSeet02,DenaLeco00}. A second, perhaps more common
mechanism is stress-induced mutagenesis; natural isolates of \textit{E.
coli} often experience an increase in their mutation rate in response to
stress\ct{BjedTena03}. As a result of these and other observations,
researchers have argued that bacterial populations evolve primarily by
periodic acquisition of mutator phenotypes followed by adaptive sweeps
and subsequent loss of the mutator\ct{GiraRadm01,DenaLeco00,NotlSeet02}.
As we shall see, the effect of this process on synonymous codon usage is
dramatic: the expected site diversity is driven by the value of $\theta$
in the wildtype regime ($\theta_w = 2 N u_w$), but the pattern of
synonymous codon usage at a site under negative selection is driven by
the value of $\theta$ in the mutator regime ($\theta_m = 2 N u_m \gg
\theta_w$). 

As a simple example of this phenomenon, we have simulated a
Fisher-Wright model of a single locus in a population of constant size
$N=1000$.  The simulated site is subject to recurrent mutation between
``alleles" $a_1$ and $a_2$ at wildtype rate $u_w=10^{-5}$. As in Section
\ref{Wright}, the alleles $a_1$ and $a_2$ differ in fitness by $s$,
where $s$ equals the mutation rate.  Periodically, we model the fixation
of a mutator allele (or, equivalently, the stress-induced mutagenesis
across the entire population) by exogenously increasing the mutation
rate to $u_m= 10^3 \times u_w$ for 100 generations; thereafter we
(artificially) enforce a selective sweep at the site, followed by
reversion to the wildtype mutation rate.  Overall, the population
experiences the mutator regime for 5\% of the time, consistent with
observed frequencies of mutator phenotypes in the
wild\ct{GiraRadm01,OlivCant00,LeclLi96}. According to our simulations,
the average site diversity, $2x(1-x)$, at a randomly chosen time equals
$0.028$, which is close to its expected value assuming that $\theta$ is
given by $\theta_w$: $\mathbb{E}[2x(1-x)]=\theta_w=0.02$.  But the
average frequency of allele $a_1$ equals $0.611$, which is close to its
expectation assuming that $\theta$ is given by $\theta_m$:
$\mathbb{E}[x]=M(\theta_m,\theta_m) = 0.616$ (Eq.  \ref{mean}).  In
other words, the average frequency of the less volatile codon $a_1$ is
dominated by the mutator periods, but the average site heterozygosity
(and any estimate of $\theta$ based on it) is dominated by the
non-mutator periods.  

There is a simple, intuitive explanation for this result.  The average
heterozygosity at the site is low at virtually all times (except during
the brief mutator periods) because selective sweeps cause monomorphism,
followed by long periods of low $\theta$. Therefore, the effective
$\theta$ for SNP diversity is small, \textit{i.e.} close to $Nu_w$.  But
the site converges quickly towards the less volatile codon during the
mutator periods, since the rate of convergence is determined by
$s=u_m$. And the site is essentially frozen during the non-mutator
periods, since the decay rate of volatility is only $u_w$.  Therefore
the expected frequency of $a_1$ at a random time  is
primarily determined by the frequency reached during the mutator regime.
As is clear from this explanation, the expected frequency of codon $a_1$
will, in general, depend upon the stochastic scheduling of mutator
periods. For example, the site will converge towards $M(\theta_m,
\theta_m)$ provided the population experiences at least one mutator
phase of duration of order $1/u_m$ generations, within every $1/u_w$
generations.  In fact, even if the mutator phases are very brief and
infrequent, the average frequency of allele $a_1$ can greatly exceed the
value predicted by $\theta$ estimated from the average site
heterozygosity.

Although the simple model used in this section does not describe any but
the most phenomenological features of mutator alleles, it does reveal an
important general observation: the value of $\theta$ estimated from
neutral SNP data does not in general equal the effective value of
$\theta$ that determines synonymous codon usage at a site under negative
selection.  This result is of utmost importance to any discussion of the
relationship between $\theta$ and the power of volatility to detect
negative selection.

\section{Positive selection} \label{PosSel} In the sections above, we have considered
selection that opposes a change to the amino acid at a site.  This type
of negative selection induces a bias towards the less volatile codons
for the favored amino acid at a site.  However, selection sometimes
favors a change in the amino acid at a particular site. In such
situations, as we will demonstrate, a site is more likely to be occupied
by a codon of greater than average volatility.

A variety of mechanisms are known to cause positive selection.
Frequency dependence often induces diversifying selection at a site,
whereas an exogenous change in the environment can induce directional
selection for a new, specific amino acid. We do not here model all of
the various types of positive selection, but rather focus on the
essential aspect shared by these mechanisms. We analyze the dynamics at
a site that has, for a period of time, experienced negative selection
for amino acid $A$, and that subsequently experiences negative selection
for different amino acid, $B$ (for whatever reason).  We refer to the
change in the selective regime as a positive selection event.

Prior to the onset of positive selection, amino acid $A$ is assigned
fitness 1 and all other amino acids fitness $1-\sigma$; subsequently,
amino acid $B$ is assigned fitness 1 and all others fitness $1-\sigma$.
We assume that $N \sigma \gg 1$ (otherwise, the site is effectively
neutral at the amino acid level) and that $\sigma \gg u$ (otherwise, the
expected codon frequencies are uniform).  Once the population shifts to
the new amino acid $B$, it is clear that the site will more likely
contain a codon that is more volatile than the average $B$-codon,
because it has just arisen through a nonsynonymous mutation. Since $B$
is now favored, negative selection subsequently operates to reduce the
volatility at the site. However, this process takes time. Thus, for some
time after the positive selection event, there is a bias toward elevated
volatility at the site, which gradually decays. In this section, we
analyze this process.

Analagously to previous sections, we initially consider a simplified
genetic code consisting of four codons, $a_1$, $a_2$, $b_1$, and $b_2$,
the first two of which encode amino acid $A$, and the latter two amino
acid $B$.  Mutations can only occur between codons $a_1$ and $a_2$,
$a_2$ and $b_2$, and $b_2$ and $b_1$, creating the mutation structure \[
a_1 \rightleftarrows a_2 \rightleftarrows b_2 \rightleftarrows b_1.  \]
In this simplified genetic code, codons $a_2$ and $b_2$ are the more
volatile codons for their respective amino acids.

After the change in selection from amino acid $A$ to $B$, a mutation to
codon $b_2$ that survives stochastic drift will eventually arise.  Thus,
at least initially, the more volatile codon $b_2$ is more prevalent
than the less volatile codon $b_1$. During this period, we can detect
the signature of the positively selected sweep because of the elevated
volatility at the site.  However, negative selection for amino acid $B$
will eventually favor codon $b_1$.  Therefore, the volatility signature
of the positive selection event will be present provided that the time
scale of decay toward codon $b_1$ is longer than the interval since the
positive selection event.

Fortunately, the time scale of decay towards $b_1$ is quite long.  For
$\theta \gg 1$, we can use the infinite population model to find this
time scale.  As discussed above, the time required to reduce the
volatility $e$-fold is of order $\frac{1}{u}$.  For $\theta \ll 1$, we
must use a finite population size calculation.  In this regime, the
population is nearly monomorphic at almost all times.  Following the
selective sweep, the site will be monomorphic for $b_2$ with almost unit
probability.  We are interested in the duration of time required such
that probability of being monomorphic for $b_2$ (as opposed to $b_1$)
has been reduced $e$-fold.  The probability of switching between $b_2$
and $b_1$, however, is of order $u$ per unit time (even before $b_2$ has
finished outcompeting $a_2$), according to Eq.  \ref{Pfix}.  Thus, the
time scale of decay in a finite population is also $\frac{1}{u}$. 

According to this analysis, a selective sweep will result in the
presence of a more volatile codon for of order $\frac{1}{u}$ generations --
a very long time indeed. (In the case of \textit{E. coli}, for example,
$\frac{1}{u}$ generations is nearly $100,000$ years, given $u \approx
5\times10^{-10}$ and generation time $\approx$ $20$ minutes. The
generation length and resulting time scale for \textit{E. coli} in the
wild may be much longer yet\ct{GibbKaps67}.) Equivalently, repeated
sweeps for amino acid changes at a site will result in the presence of
more volatile codons at almost all times, provided that new sweeps occur
more often than every $\frac{1}{u}$ generations.

\subsection{Inferring Positive Selection}

The above analysis for a simplified genetic system generalizes in an
obvious way to the real genetic code.  After a positive selection event
at a site, the population switches from a codon for amino acid $A$ to a
codon for amino acid $B$.  The expected volatility of the new codon is
greater than the average volatility of $B$-codons, because the new codon
has just arisen through a nonsynonymous mutation.  To be more precise,
if the population is monomorphic for a random non-$B$ codon before the
selective sweep, then after the sweep occurs the expected relative
frequencies of the $B$-codons are given, approximately, by their relative
volatilities.  Subsequent to the selective sweep, the increased
volatility at the site will decay on a time scale of order of
$\frac{1}{u}$ generations.

There is a critical distinction between the volatility signature of
positive selection versus that of negative selection.  The depressed
volatility at a site under negative selection is caused by a
mutation-selection-drift balance. When the effective population size is
small, a large number of sites are required to distinguish negative
selection from neutrality. By contrast, the volatility signature of
\textit{positive} selection is {\it not} an equilibrium property, and
it is not sensitive to population size.  Regardless of $\theta$, the
probability of sampling a more volatile codon is significantly elevated
immediately after a selective sweep at a site, and this probability
decays only at rate $u$.

As we have seen, a gene that contains many sites under positive
selection will exhibit a greater volatility (controlling for its amino
acid composition) than a gene under neutral or, especially, negative
selection.  How many positively selected sites are required in order to
detect a reliable signal? In the case of Leucine, for example, the
expected volatility of a site that has recently experienced a positively
selected sweep is approximately $0.660 \pm 0.072$ (one standard
deviation), whereas a neutral Leucine site has expected volatility
$0.646 \pm 0.073$, and a Leucine site under negative selection has
expected volatility $0.632 \pm 0.0070$ (see Table \ref{GLRS}).
Therefore, the volatility of about 100 Leucine sites under positive
selection will be significantly greater (at the 95\% confidence level)
than that of 100 neutral sites.  Similarly, the volatility of about 25
positively selected Leucine sites will be significantly greater than that of
25 negatively selected sites.  Similar results hold for Serine and
Arginine; Glycine is less informative.

It is worth noting that the elevated volatility for a positively selected
Serine site will decay even more slowly than for other amino acids,
because the highly volatile codons ACC and AGT are not connected by
synonymous mutations to other serine codons. 

\section{Discussion}
\label{Discussion}

\subsection{Codon volatility versus comparative sequence analysis}

Selection pressures on proteins are usually estimated by comparing
homologous nucleotide sequences\ct{ZuckPaul65}.  Orthologous genes are
identified in different organisms and sequenced; their sequences are
then aligned, and the changes that have accumulated since divergence are
used to infer the selection pressures that have been
acting\ct{GoldYang94}. When available, sequence variation sampled from
individuals within a species can be compared with variation across
species to produce an elegant test for adaptive evolution at a
locus\ct{McDoKrei91,SawyHart92}. In addition, there are a variety of
statistical tests designed to detect a departure from neutrality in the
site frequency spectrum sampled within a single species (see ref.\ct{Krei00}
and references therein). In many cases, the complete distribution of
these statistics under the neutral null model are difficult to derive,
but they have been studied through computer simulation\ct{SimoChur95}.

Techniques for estimating selective constraints via sequence comparison
are typically applied, independently, to one or several genes at a time.
When extensive intra- or inter-specific sequence data are available at a
locus of interest, such techniques have proven enormously useful for
measuring selection, and it is unlikely that they will be significantly
improved by incorporating information about synonymous codon usage.  But
the accurate estimation of selective constraints requires a large number
(approximately six or more\ct{AnisBiel01}) of orthologous sequences for
each gene of interest.  At the genome-wide scale, comparative data
(\textit{i.e.} orthologous gene sequences) will not be available for all
genes, and methods to estimate selective constraints based on sequence
comparison will often be inapplicable.  Furthermore, the genes under
positive selection are often of particular interest, but such genes are
even less likely to have identifiable orthologs in related species due
to their rapid sequence divergence\ct{PlotDush04}.  Even in the lineage
of the \textit{Saccharomyces} genus, which is currently the best-case
scenario for comparative genomics, the genomes of four species have been
fully sequenced and only two-thirds of the genes in \textit{S.
cerevisiae} have unambiguously identifiable orthologs in related
species\ct{PlotFras04}. Unlike comparative techniques, the analysis of
synonymous codon usage offers a computational tool to screen for
selection pressures on \textit{all} genes in a sequenced genome. Genome-wide
screens based on analyzing synonymous codon usage may prove useful in
identifying important classes of genes under strong selection, such as
the antigens of pathogens\ct{PlotDush04}.

Unlike most comparative statistics that test for a departure from
neutrality, estimates of selection based on bootstrapped volatility
scores\ct{PlotDush04} are not `estimators' in a rigorous statistical
sense -- \textit{i.e.} statistics whose sampling properties can be
derived from a null model, and which can be used in likelihood ratio
tests of a null hypothesis\ct{YangNiel00,ClarkGlan03}. Given the
expected relative frequencies of codons that we have derived for each of
the three regimes (neutral, negative, and positive selection; Table
\ref{GLRS}), it may yet be possible to design maximum-likelihood methods
that estimate the number of sites of a gene in each regime. This
approach will be complicated, however, by other sources of codon bias;
see below.

Aside from the different situations in which they are applicable, and
differences in the rigor of their derivation, estimates of selection
based on codon volatility differ in a fundamental way from most
estimates based on sequence comparison.  Homologous sequence comparison
between species is often used to assess, either by maximum
likelihood\ct{GoldYang94} or maximum parsimony\ct{Li93}, the rates of
synonymous and non-synonymous substitutions in a coding sequence. The
ratio of these rates, dN/dS, is used as a measure of the selective
constraints that have been acting on a protein since the divergence of
the species being compared.  An alternative approach, based on a Poisson
Random Field (PRF) model of mutation frequencies, uses the site
frequency spectrum at a locus sampled from individuals within a species
to deduce the average selective pressure for or against amino acid
changes in a gene\ct{SawyHart92}. (Poisson Random Field models can also
be used to construct likelihood ratio tests of departure from
neutrality\ct{BustWak01}.) Like most comparative methods, however, both
of these models typically assume that all the sites within a gene
experience the same selective pressure against amino acid substitutions
(but see the site-by-site likelihood tests of Yang \textit{et
al.}\ct{YangNiel00}).  Under the PRF theory, for example, authors have
estimated a very small ``average'' selection pressure against amino acid
changes in \textit{E.  coli} genes: $\sigma \sim
10^{-8}$\ct{HartSawy94}.  This value does not represent the arithmetic
average of the true $\sigma$ values across sites, but rather the
best-fit constant value of $\sigma$ that would make the PRF model
consistent with observed sequence variation at polymorphic sites.

When evolutionary rates are estimated at \textit{individual}
residues\ct{Yang00,YangNiel00}, however, we find great variation across
sites. Moreover, direct experimental measurements of the fitness
consequences of amino acid substitutions in micro-organisms reveals huge
variation in selection pressures across the residues of an individual
protein: a substantial proportion of substitutions are lethal, and a
substantial proportion have undetectable
effect\ct{WertDrub92,WlocSzaf01,ZeylDeVi01,SanjMoya04}. Therefore, it is
not entirely clear how best to interpret the value of $\sigma \sim
10^{-8}$ estimated for \textit{E. coli} genes using the PRF model, which
assumes constant pressure at each residue. 

Compared to dN/dS or $\sigma$ estimated by the PRF model, codon
volatility quantifies selection pressures in a very different, coarser
manner.  As discussed above, volatility essentially measures the number
of sites in a gene that experience negative ($\sigma \gg u$) versus
neutral ($\sigma \ll u$) versus positive selection. Given that, in
reality, many amino acid changes to a protein sequence are lethal while
other changes have no effect whatsoever, it is reasonable and meaningful
to estimate the number of sites in the selected versus neutral regimes.
But volatility is not sensitive to variation in selective pressures
within either of these regimes. Hence, the volatility measure is in some
ways a coarser description of selective pressure than PRF or dN/dS.  One
should not necessarily expect that volatility will correlate very
strongly with dN/dS or PRF estimates, because the latter measures
represent some sort of average $\sigma$ over the entire gene, and are
thus presumably sensitive to the full range of variation in $\sigma$.  A
measure based on codon volatility is therefore different from and
complementary to dN/dS or PRF estimates of the selective
constraints on a genes.

As an aside, it is important to note that the most common model used to
estimate dN/dS from divergent nucleotide sequences\ct{GoldYang94} does
not itself reflect the relationship between selection and volatility.
dN/dS is often estimated by fitting maximum likelihood parameters to a
simplified Markov-chain model of sequence evolution that ignores
population variability\ct{GoldYang94}.  Models that ignore population
variability are perfectly reasonable approximations when comparing the
sequences of relatively divergent lineages\ct{GoldYang94}; but such
models fail to detect the effect of amino-acid selection on synonymous
codon usage.  Such models consider only a single sequence that is
assumed to represent the dominant genotype in the population at any
time.  Mutation and selection are modeled simultaneously by adjusting
the transition rates between codon states in the
sequence\ct{GoldYang94}.  As a result, in equilibrium, the number of
transitions into a state per unit time must equal the number of
transitions out of that state; and so equilibrium synonymous codon usage
does not depend upon the strength of selection in these simplified
models\ct{GoldYang94}. (In fact, under the standard assumption of
time-reversibility, such models require as parameters the specification
of the equilibrium codon usage\ct{GoldYang94}, and therefore they
clearly cannot be used to predict equilibrium codon usage.) Simulations
of sequence evolution based on these simplified models (such as the
non-frequency-dependent simulations of Zhang\ct{Zhan04}) will thus fail
to detect the relationship between dN/dS and volatility, whereas more
detailed simulations that account for population variability (such as
the frequency-dependent simulations of Zhang\ct{Zhan04}, as well as the
non-frequency-dependent simulations in this work) will properly reflect
the relationship between selection and volatility, as predicted by
Fisher-Wright models of a replicating population.

\subsection{Other sources of codon bias} 

Although it came as a surprise to early neutral
theorists\ct{KingJuke69}, it is now clear that there are several
processes that result in unequal usage of synonymous codons.  Many
processes that cause codon bias in microorganisms, such as biased
nucleotide content or mutation rates, can apply roughly equally to all
the genes in a genome.  To the extent that other sources of codon bias
apply equally across a genome, it is straightforward to control for
these biases when comparing the volatilities of genes within a genome to
estimate selection pressures on proteins\ct{PlotDush04}.

To the extent that other sources of codon bias differ from gene to gene
within a genome, they may (if not properly controlled for) introduce
errors into estimates of the relative selection pressures on proteins
inferred from codon volatility\ct{PlotDush04}. Similarly, selection on
synonymous codons -- particularly selection that varies from gene to
gene -- will likewise introduce errors into estimates of selection on
protein sequences obtained by comparative techniques such as
dN/dS\ct{SharLi87,HirsFras04}.

As we have argued, some of the variation in synonymous codon usage
across a genome is caused by the variation in selection pressures on
protein sequences.  Throughout our analysis, we have specifically
ignored any other source of codon biases so as to derive the effects of
selection at the amino acid level on codon usage.  But in many organisms
other processes that vary between genes are certainly operating as well.
For instance, it is known that the transition/transversion mutation bias
can vary across a genome.  Results on \textit{S. cerevisiae}, whose
genome exhibits marked variation in the tr/tv bias\ct{PlotFras04},
suggest that this source of variable codon bias will not distort
estimates of selection based on volatility: whether or not one accounts
for the variation in the tr/tv bias across the genome of \textit{S.
cerevisiae} one obtains virtually the same rankings of gene volatilities
($r>0.99$)\ct{PlotFras04}. 

Aside from mutational biases, there are other sources of codon bias that
vary from gene to gene in some organisms. In the yeast \textit{S.
cerevisiae}, researchers have observed that synonymous codon usage,
measured by the Codon Adaptation Index (CAI)\ct{SharLi87}, is correlated
with a gene's expression level in laboratory conditions\ct{CoghWolf00}.
This correlation is thought to be caused by selection for translational
efficiency and/or accuracy: a codon corresponding to a more abundant tRNA is
expected to be translated more quickly (due to the higher probability
per unit time that the appropriate tRNA will ``find" the codon) and more
accurately (since the correct tRNA will likely have the greatest chance
of pairing if it is the most abundant). 

Considering this alternative source of biased codon usage, two questions
should be asked: do other sources of codon bias distort estimates of
selection based on volatility, and how can we control for these
confounding factors? Unfortunately we do not have a truly satisfactory
answer for either of these questions, but the discussion below may shed
some light on the issues involved.  

With regards to the first question, we note that the degree to which
other sources of codon bias may distort volatility-based estimates of
selection will strongly depend on the organism being studied.  Some
species (such as humans) exhibit a much weaker correspondence between
codon frequencies and tRNA abundances than others species; so clearly
other sources of codon bias will affect volatility values differently in
different species. In a species with a strong correspondence between
codon usage and tRNA abundances, the extent to which variation in this
source of codon bias across the genome affects volatility will depend on
whether volatile codons are (un)preferred: if there is no correlation
between volatility and tRNA abundances, then the other sources of codon
bias will only introduce random error into volatility estimates, making
them less powerful but still reliable. If instead the preferred codons
tend to have either high or low volatility, then this effect could
introduce systematic errors into volatility estimates. In the latter
case, in order to quantify how much codon usage bias is caused by
volatility as opposed to other factors, one would require a method to
predict for individual genes the amount of codon bias due to these other
factors. Unfortunately we are far from having the necessary level of
predictive power for other sources of codon bias in any organism.
Although gene expression level is somewhat predictive of codon bias,
expression levels do not explain most of the variation in codon bias in
any genome studied thus far\ct{Akas01,CoghWolf00}.  Until the various
sources of biased codon usage can be reliably disentangled, we cannot
reliably quantify the effects of these biases on volatility-based
estimates of selection.

The second question, how to control for other sources of biased codon
usage, is also difficult to answer at present.  As discussed above, an
appropriate method to control for other sources of bias would require
disentangling the various sources of codon bias in a predictive manner
for each gene. While this degree of precision is not currently possible,
one approach is to assume that the codon bias measured by CAI is
entirely independent of volatility, and then control for CAI using
partial correlations. For several reasons, we expect this approach to be
conservative, as we illustrate using the yeast \textit{S. cerevisiae}
(we use this species as an example because it shows a strong preference
for codons that match abundant tRNAs, and because we have reliable dN/dS
values for almost two-thirds of its genes, calculated from multiple
alignments of closely related species\ct{HirsFras04}). First, we note
that dN/dS is itself strongly correlated with both CAI and gene
expression levels\ct{PalPapp01}, and it is therefore impossible to
construct any measure of selective constraint that agrees with dN/dS and
is not itself strongly correlated with CAI and expression levels in
yeast. Second, it is possible that the codon bias measured by CAI is in
part \textit{caused} by volatility (\textit{i.e.} highly expressed genes
tend to experience stronger purifying selection and therefore exhibit
unusual codon usage biased towards lower volatility), and so controlling
for CAI would be inappropriate. Despite several biological hypotheses,
there is no accepted mechanistic explanation for the correlation between
CAI and dN/dS in yeast\ct{PalPapp01, Akas01}, and so it is unclear
whether controlling for CAI is appropriate.
Nevertheless, we have tested the correlation between volatility and
dN/dS while controlling for CAI using a partial correlation. We find
that even when controlling for CAI (or expression levels), there remains
a highly significant correlation between volatility and dN/dS in yeast
($p<10^{-34}$\ct{PlotFras04}).  Therefore, even under this conservative
test, estimates of selection obtained by volatility are still consistent
with estimates obtained by homologous sequence comparison. We interpret
this result as evidence that volatility is measuring selective
constraints above and beyond any signal inherent in CAI.

Indeed, there is a great deal of empirical evidence
indicating that the volatility of a gene is correlated with the
selective constraint it experiences.  Aside from highly significant
correlations between volatility and dN/dS in bacterial species and
yeast\ct{PlotDush04}, volatility also reflects a range of other features
known to correlate with selection on proteins. In \textit{S.
cerevisiae}, for example, volatility is strongly correlated with the
essentiality of genes, the number of their protein-protein interactions,
and the degree to which they are preserved throughout the eukaryotic
kingdom\ct{PlotFras04}.  Furthermore, volatility is significantly
elevated among the known antigens and surface proteins (which experience
positive selection) in the pathogens \textit{Mycobacterium
tuberculosis}, \textit{Plasmodium falciparum}, and Influenza A
virus\ct{PlotDush03,PlotDush04}. And volatility is significantly
depressed in the genes essential for growth of \textit{M.
tuberculosis}, as well as in the genes conserved between related
\textit{Mycobacterium} species\ct{PlotDush04}. Therefore, despite
potential confounding sources of codon bias that cannot at present be
controlled for with appropriate accuracy, in practice volatility-based
methods produce estimates of selection pressures that are consistent
with our understanding of protein evolution over a diverse range of
taxa.

Finally, we note that there may be direct selection on synonymous codons
in order to evade mistranslation\ct{Kono85}. Since mistranslation is far
more likely to occur between a codon and an anticodon that differ by a
single nucleotide, the definition of volatility (Eq. \ref{voldef})  is
appropriate for measuring the selective pressure for or against 
mistranslation. The strength of this type of selection on synonymous
codons would depend upon the mis-incorporation rate of tRNA (which is
far higher than the mutation rate) and the detriment of mistranslation
(which is likely far lower than that of most mis-sense mutations). It is
difficult at present to measure the molecular parameters of tRNA
mis-incorporation and its fitness effects; so it is unclear how much of
a volatility signal arises from mistranslation avoidance versus standard
selection on mis-sense mutations. However strong this signal, though,
the volatility of a gene would still reflect the degree to which there
is selection to conserve, or not to conserve, the (translated) protein
sequence.

\section*{Acknowledgments}

We thank Daniel Fisher, Andrew Murray, and Michael Turelli for their
input during the preparation of this manuscript. J.B.P. acknowledges
support from the Harvard Society of Fellows. M.M.D. acknowledges
support from a Merck Award for Genome-Related Research.

\bibliography{./bib}

\end{document}